\documentclass[3p,times,procedia]{elsarticle}
\usepackage{nupha_ecrc}

\volume{00}

\firstpage{1}

\journalname{Nuclear Physics A}

\runauth{}

\jid{nupha}

\jnltitlelogo{Nuclear Physics A}

\usepackage{amssymb}

\usepackage[figuresright]{rotating}

\usepackage{amsmath}
\usepackage{xspace}
\usepackage{graphicx}

\newcommand{\Npart}{\mbox{$N_{\rm part}$}\xspace}

\newcommand{\Et}{\mbox{${\rm E}_T$}\xspace}

\newcommand{\sqsn}{\mbox{$\sqrt{s_{_{NN}}}$}\xspace}

\newcommand{\pp}{\mbox{$p$$+$$p$}\xspace}
\newcommand{\dau}{\mbox{$d$$+$Au}\xspace}

\newcommand{\auau}{\mbox{Au$+$Au}\xspace}
\newcommand{\pbpb}{\mbox{Pb$+$Pb}\xspace}
\newcommand{\cucu}{\mbox{Cu$+$Cu}\xspace}
\newcommand{\cuau}{\mbox{Cu$+$Au}\xspace}
\newcommand{\uu}{\mbox{U$+$U}\xspace}
\newcommand{\heau}{\mbox{$^{3}$He$+$Au}\xspace}

\newcommand{\Nqp}{\mbox{$N_{\rm qp}$}\xspace}
\newcommand{\dEt}{\mbox{$d{\rm E}_T/d\eta$}\xspace}

\newcommand{\dEtNorm}{\mbox{$(d{\rm E}_T/d\eta) / (0.5 \Npart)$}\xspace}

\newcommand{\dEtNormQ}{\mbox{$(d{\rm E}_T/d\eta) / (0.5 \Nqp)$}\xspace}

\newcommand{\ebj}{\mbox{$\varepsilon_{BJ}$}\xspace}

\begin{document}

\begin{frontmatter}



\dochead{}

\title{Transverse Energy Measurements from the Beam Energy Scan in PHENIX}


\author{J.T. Mitchell (for the PHENIX Collaboration)\footnote{For the full PHENIX Collaboration author list and acknowledgements, see Appendix ``Collaboration''  of this volume.}}

\address{Brookhaven National Laboratory, Building 510C, P.O. Box 5000, Upton, NY 11973-5000}

\begin{abstract}
Transverse energy distributions at midrapidity have been measured by the PHENIX experiment at the BNL Relativistic Heavy Ion Collider (RHIC) for \auau, \uu, \cuau, \cucu, \heau, \dau, and \pp collisions over a wide energy range from \sqsn = 7.7 GeV to \sqsn = 200 GeV as a function of centrality.  For central \auau collisions, it is observed that the midrapidity Bjorken energy density demonstrates a power law behavior from \sqsn = 7.7 GeV to \sqsn = 2.76 TeV. At a given collision energy, the data presented as a function of \Npart are independent of the size of the collision system. For \auau, \cuau, and \cucu collisions, the centrality-dependent data are better described by scaling with the number of constituent quark participants than scaling with the number of nucleon participants.
\end{abstract}


\end{frontmatter}


\section{Introduction}
\label{sec:intro}
The PHENIX experiment at the BNL Relativistic Heavy Ion Collider (RHIC) has compiled a comprehensive dataset from the years 2000 to 2015 that includes collisions of \auau, \uu, \cuau, \cucu, \heau, \dau, and \pp collisions at a variety of collision energies.  This dataset can be exploited to study the dynamics of the colliding system with measurements of transverse energy production at midrapidity, \dEt, as a function of collision energy and centrality.  This study is complementary and extends previous studies of dynamics based on charged particle multiplicity production by the PHOBOS experiment \cite{Alver:2010ck}. Although PHENIX has also measured charged particle multiplicity \cite{Adler:2004zn, Adare:2015bua}, only transverse energy measurements will be presented in this article. More details on the PHENIX transverse energy measurements presented here can be found elsewhere \cite{Adler:2004zn, Adler:2013aqf, Adare:2015bua}.

\section{Energy Dependence of Transverse Energy Production}
\label{sec:ebj}

\begin{figure}[htbp]
\begin{center}
\includegraphics[width=0.4\textwidth]{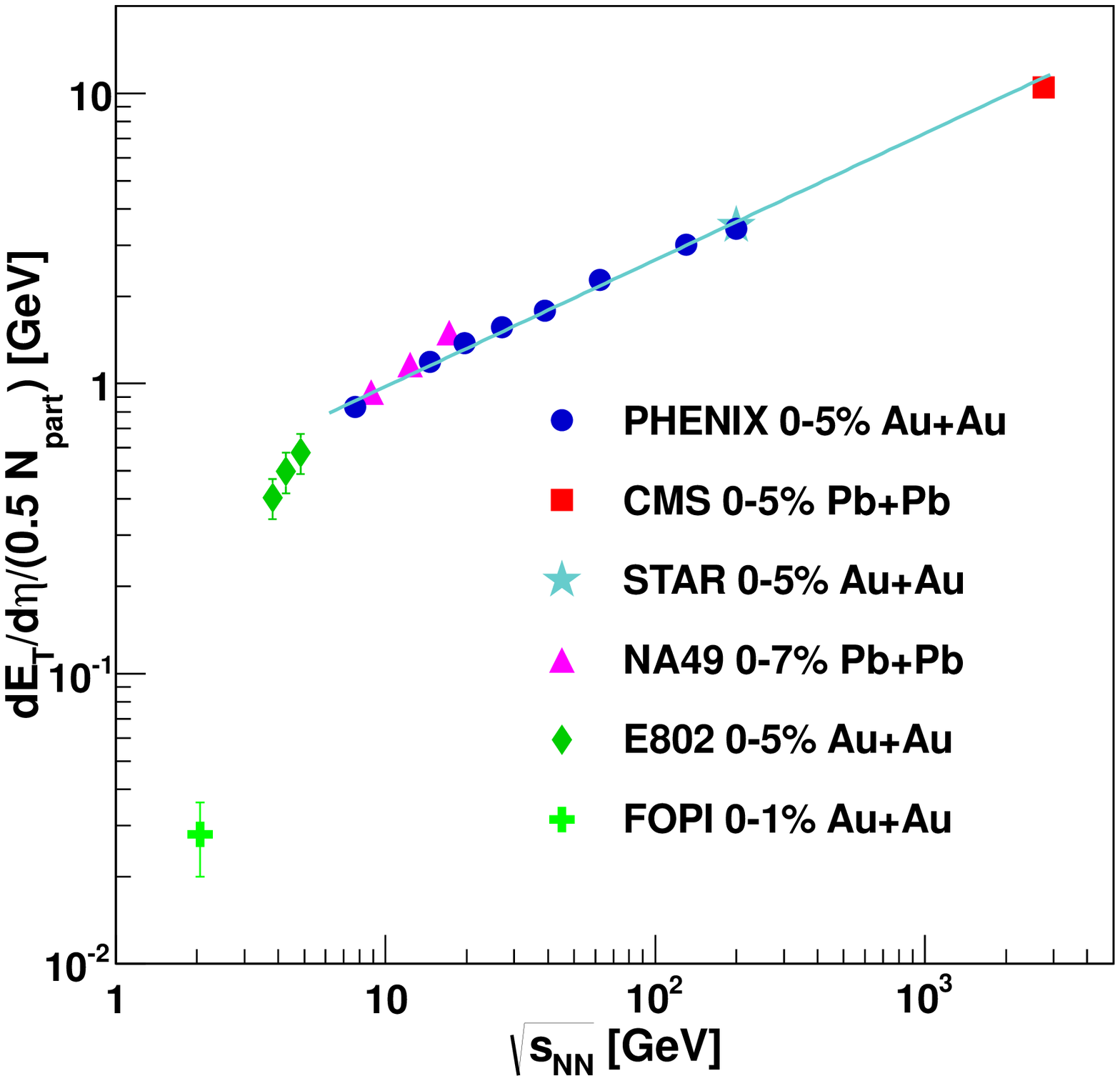}
\includegraphics[width=0.4\textwidth]{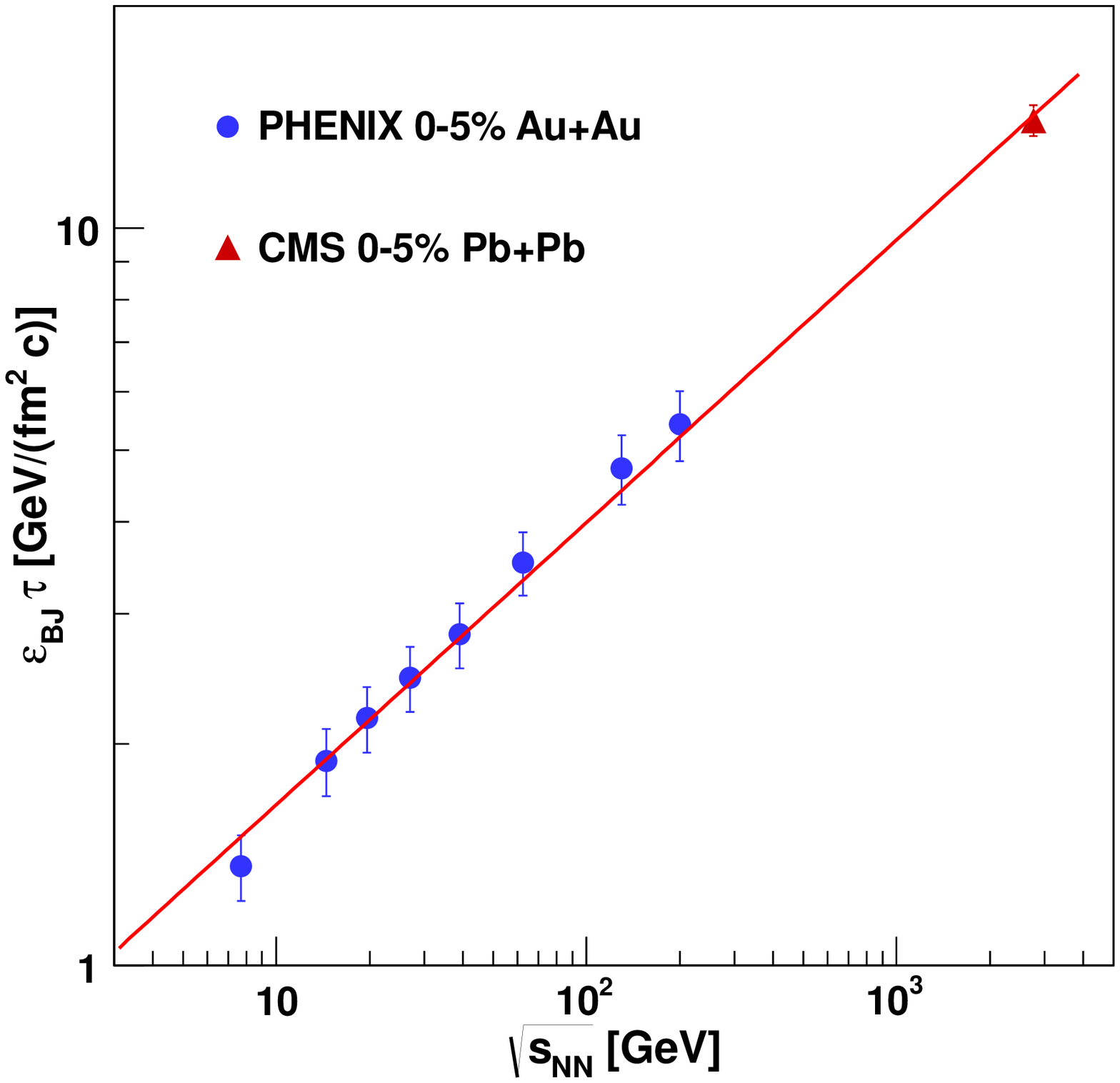}
\end{center}
\caption{(Left) \dEtNorm as a function of \sqsn for central \auau and \pbpb collisions. In addition to the PHENIX data, data are shown from FOPI \cite{Reisdorf:1996qj}, E802 \cite{Ahle:1999jm}, NA49 \cite{Afanasiev:2002fk}, STAR \cite{Adams:2004cb}, and CMS \cite{Chatrchyan:2012mb}. The line is a power law fit to the PHENIX data. (Right) \ebj multiplied by $\tau$ as a function of \sqsn for central \auau and \pbpb collisions. In addition to the PHENIX data, data are shown from CMS \cite{Chatrchyan:2012mb}. The line is a power law fit to all of the data points.}
\label{fig:etExcite}
\end{figure}

Figure \ref{fig:etExcite} (left) shows transverse energy production, \dEtNorm, for central \auau and \pbpb collisions at midrapidity as a function of \sqsn.  Above \sqsn = 7.7 GeV, the value of \dEtNorm is observed to be well described by a power law, $\dEtNorm \propto \sqsn^{b}$, where the exponent is $b = 0.428 \pm 0.021$.  These results can also be expressed in terms of the Bjorken energy density \cite{Bjorken:1982qr}
\begin{equation}
  \varepsilon_{BJ} = \frac{1}{A_\perp \tau} \frac{d\Et}{d\eta}
\end{equation}
where $A_\perp$ is the transverse overlap area of the nuclei determined from the Glauber model and $\tau$ is the formation time, typically estimated to be 1 fm/c. Figure \ref{fig:etExcite} (right) shows the Bjorken energy density multiplied by the formation time for central \auau and \pbpb collisions above \sqsn = 7.7 GeV. These data are well described by $\varepsilon_{BJ} \tau \propto \sqsn^{b}$, where $b = 0.422 \pm 0.035$.

\section{System Size Dependence of Transverse Energy Production}
\label{sec:size}

\begin{figure}[!h]
\begin{center}
\includegraphics[width=0.5\textwidth]{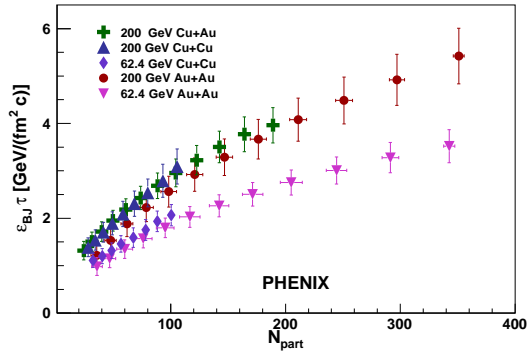}
\end{center}
\caption{\ebj as a function of \sqsn for \auau, \cuau, and \cucu collisions at \sqsn = 200 and 62.4 GeV.  The error bars represent the statistical and systematic errors.}
\label{fig:ebjCu}
\end{figure}

The PHENIX dataset includes \auau, \cuau, and \cucu collisions at \sqsn = 200 GeV along with \auau and \cucu collisions at \sqsn = 62.4 GeV. This facilitates a study of the system size dependence of transverse energy production.  Figure \ref{fig:ebjCu} shows the Bjorken energy density multiplied by the formation time as a function of \Npart for these systems. At a given collision energy, \ebj for systems of differing sizes are consistent with each other.  This demonstrates that \ebj is independent of the system size at \sqsn = 200 and 62.4 GeV.

\section{Centrality Dependence of Transverse Energy Production}
\label{sec:centrality}

\begin{figure}[htbp]
\begin{center}
\includegraphics[width=0.4\textwidth]{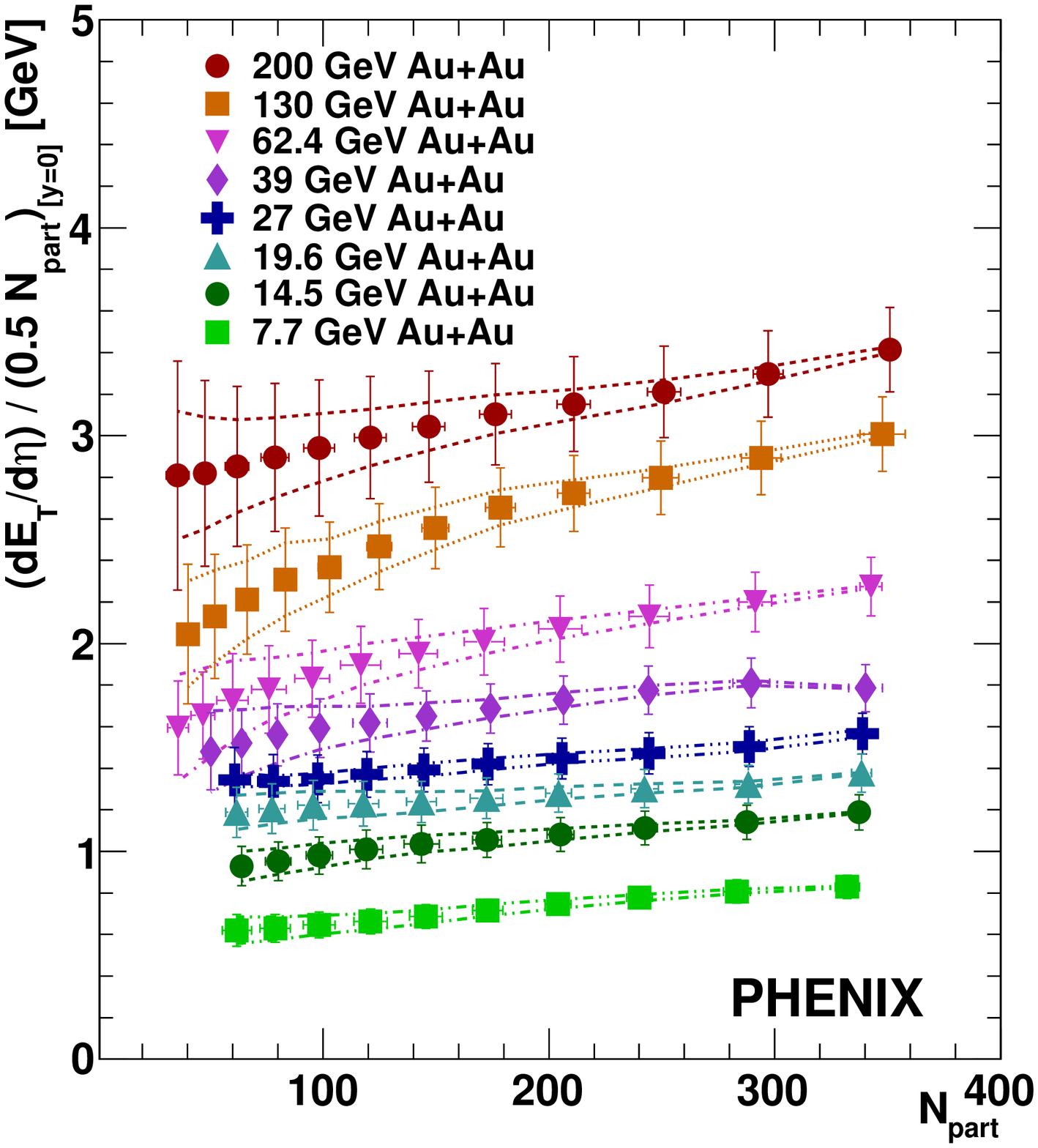}
\includegraphics[width=0.4\textwidth]{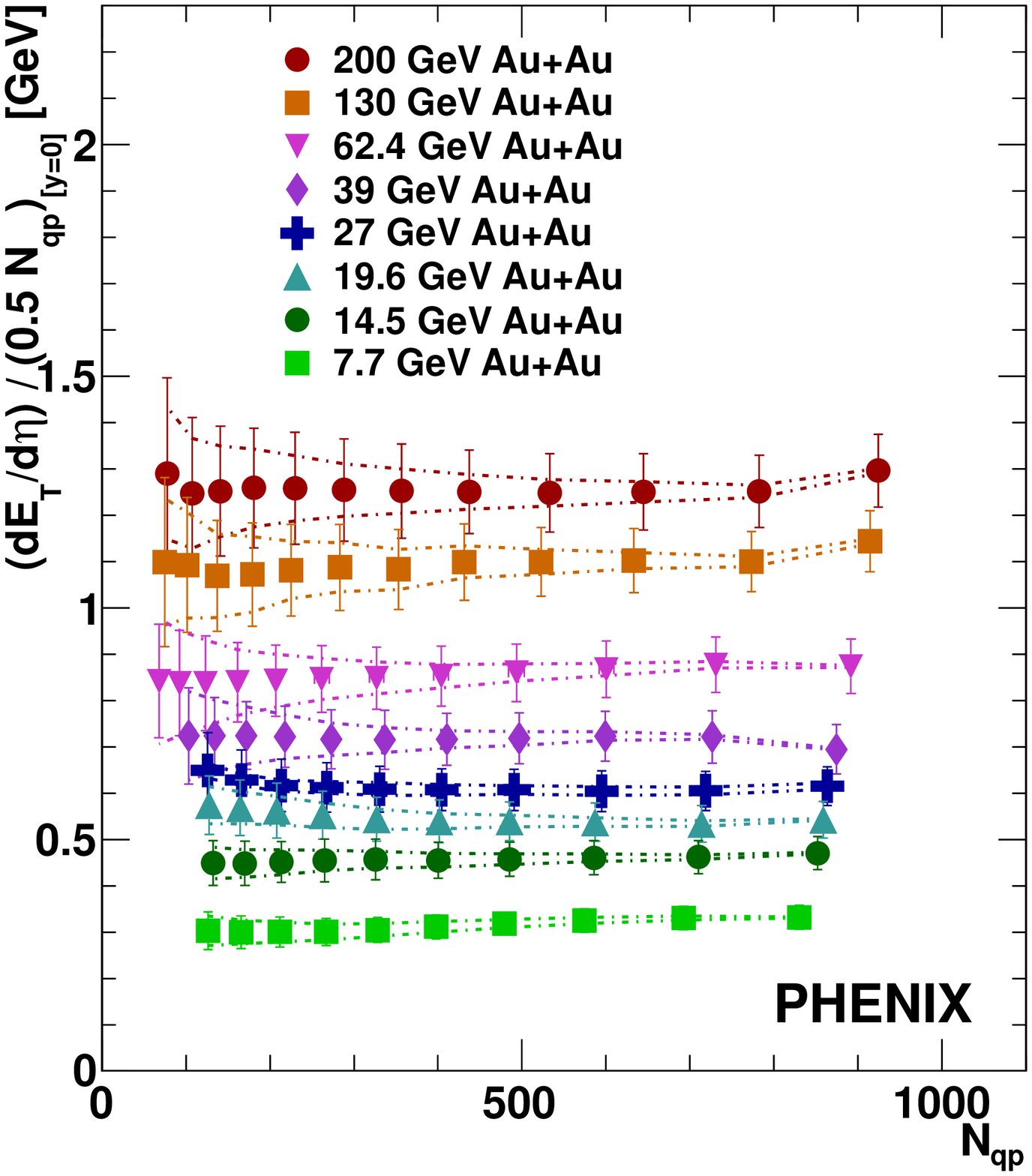}
\end{center}
\caption{(Left) \dEtNorm as a function of \Npart for \auau collisions from \sqsn = 200 GeV to 7.7 GeV. (Right) \dEtNormQ as a function of \Nqp for \auau collisions from \sqsn = 200 GeV to 7.7 GeV. For both plots, the lines bounding the points represent the trigger efficiency uncertainty within which the points can be tilted. The error bars represent the remaining statistical and systematic errors.}
\label{fig:etNpart}
\end{figure}

\begin{figure}[!h]
\begin{center}
\includegraphics[width=0.4\textwidth]{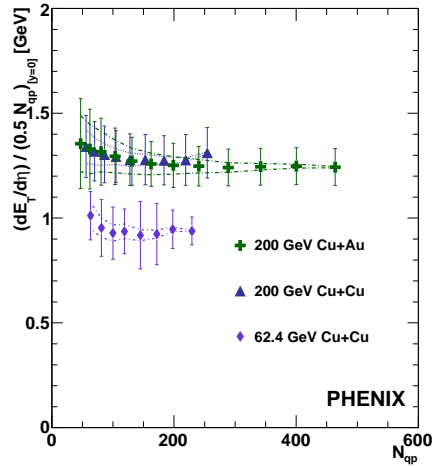}
\end{center}
\caption{\dEtNormQ as a function of \Nqp for \cuau and \cucu collisions for \sqsn = 200 GeV and \sqsn = 62.4 GeV. The lines bounding the points represent the trigger efficiency uncertainty within which the points can be tilted. The error bars represent the remaining statistical and systematic errors.}
\label{fig:etNqpCu}
\end{figure}

The centrality dependence of \dEtNorm is typically expressed in terms of the number of nucleon participants, \Npart, as shown in Figure \ref{fig:etNpart} (left) for \auau collisions from \sqsn = 7.7 to 200 GeV. Note that the midrapidity data increase with increasing \Npart and are not consistent with scaling by the number of nucleon participants in \auau collisions from \sqsn = 200 GeV all the way down to \sqsn = 7.7 GeV.  The data can also be examined as a function of centrality expressed as the number of constituent quark participants, \Nqp \cite{Eremin:2003qn}.  This has been estimated using a Glauber model calculation that has been modified to replace nucleons with their constituent quarks \cite{Adare:2015bua}. The results are shown in Figure \ref{fig:etNpart} (right), which shows \dEtNormQ as a function of \Nqp for \auau collisions from \sqsn = 200 GeV down to 7.7 GeV.  For all energies, the data are better described by scaling with \Nqp than scaling with \Npart.  This is consistent with previous studies of charged particle multiplicity distributions measured by PHOBOS down to \sqsn = 19.6 GeV \cite{Nouicer:2006pr}. Figure \ref{fig:etNqpCu} shows \dEtNormQ as a function of \Nqp for \cuau and \cucu collisions at \sqsn = 200 and 62.4 GeV.  The data are also better described by scaling with \Nqp for the smaller systems.

\section{Summary}
\label{sec:summary}

The PHENIX experiment has completed a systematic survey of transverse energy production in a variety of collision systems for \sqsn = 7.7 GeV to \sqsn = 200 GeV.  The Bjorken energy density for central \auau collisions at midrapidity is well described by a power law from \sqsn = 7.7 GeV up to \sqsn = 2.76 TeV. At \sqsn = 200 and 62.4 GeV, \dEtNorm presented as a function of \Npart is independent of the size of the system.  For \sqsn = 7.7 to 200 GeV, it is observed that transverse energy production is better described by scaling with respect to the number of constituent quark participants than by the number of nucleon participants.





\bibliographystyle{elsarticle-num}
\bibliography{qm2015-mitchell-bbl}







\end{document}